\documentclass[12pt]{article}
\pdfoutput=1
\usepackage{putex}
\usepackage{feyn}
\usepackage[vcentermath]{youngtab}
\usepackage{graphicx}
\usepackage{epstopdf}
\usepackage{enumerate}
\usepackage{cite}
\usepackage{tensor}
\usepackage{slashed}
\usepackage{feynmf}
\usepackage{hyperref}

\numberwithin{equation}{section}

\renewcommand{\baselinestretch}{1.2}

\def\sc{{\rm sc}}
\def\vac{{\rm vac}}
\def\d{\delta}

\def\bul{$\bullet$~}
\def\Oc{{\cal O}}
\def\vs{\vskip .1 in}
\def\D{\Delta}
\def\hyp{{}_2F_1}

\def\L{\Lambda}
\def\gmn{\gamma_{mn}}
\def\g{\gamma}
\def\a{\alpha}

\newcommand{\e}[2] {\begin{equation} \label{#1} #2 \end{equation}}
\def\gmn{\gamma_{mn}}
\def\rmn{R_{mn}}
\def\pmn{P_{mn}}
\def\amn{A_{mn}}

\def\gsc{\gmn^{\rm sc}}
\def\cFsc{{\cal F}^{\rm sc}}
\def\csc{\chi^{\rm sc}}

\newcommand {\be} {\begin {equation}}
\newcommand {\ee} {\end {equation}}

\newcommand {\bes} {\begin {equation*}}
\newcommand {\ees} {\end {equation*}}

\newcommand{\es}[2] {\begin{equation} \label{#1} \begin{split} #2 \end{split} \end{equation}}

\newcommand{\Z}{\mathbb{Z}}
\newcommand{\N}{\mathbb{N}}

\newcommand{\C}{\mathbb{C}}

\newcommand{\beq}{\begin{equation}}
\newcommand{\eeq}{\end{equation}}

\newcommand{\cO}{\mathcal{O}}

\newcommand{\cF}{\mathcal{F}}

\def\be{ \begin{equation} }
\def\ee{ \end{equation} }

\def\half{{1\over  2}}

\def\sec{\section}
\def\subsec{\subsection}
\def\N{{\cal N}}

\def\C{{\cal C}}
\def\cF{{\cal F}}

\def\L{\Lambda}

\def\eqr{\eqref}

\def\b{\beta}

\def\C{{\cal C}}

\newcommand{\bea}{\begin{eqnarray}}
\newcommand{\eea}{\end{eqnarray}}

\newcommand\zb{\bar{z}}

\def\rar{\rightarrow}
\def\C{\mathcal{C}}
\def\hb{\overline{h}}
\def\zb{\overline{z}}

\def\eps{\epsilon}

\begin{document}

\institution{PU}{Department of Physics, Princeton University, Princeton, NJ 08544, USA}

\title{Virasoro conformal blocks in closed form}

\authors{Eric Perlmutter}

\abstract{Virasoro conformal blocks are fixed in principle by symmetry, but a closed-form expression is unknown in the general case. In this work, we provide three closed-form expansions for the four-point Virasoro blocks on the sphere, for arbitrary operator dimensions and central charge $c$. We do so by solving known recursion relations. One representation is a sum over hypergeometric global blocks, whose coefficients we provide at arbitrary level. Another is a sum over semiclassical Virasoro blocks obtained in the limit in which two external operator dimensions scale linearly with large $c$. In both cases, the $1/c$ expansion of the Virasoro blocks is easily extracted. We discuss applications of these expansions to entanglement and thermality in conformal field theories and particle scattering in three-dimensional quantum gravity.}

\date{}

\maketitle

\tableofcontents

\renewcommand{\baselinestretch}{1.2}

\sec{Introduction}\label{i}

This paper addresses a somewhat technical matter at the heart of two-dimensional conformal field theory: the determination of the Virasoro conformal blocks for four-point correlation functions on the sphere. 

It has been known since the seminal work of \cite{bpz} that the Virasoro blocks, which contain all contributions of a given conformal family to a four-point function, are fully fixed by the local conformal symmetry in terms of the external and internal operator dimensions and central charge, $c$. Nevertheless, a complete expression for the general Virasoro blocks eludes us. It is hard to overstate the centrality of this object to 2d CFT. Furthermore, Virasoro symmetry appears in increasingly many contexts across dimensions, including the AGT correspondence \cite{Alday:2009aq}, asymptotic symmetries of quantum gravity in four-dimensional Minkowski space \cite{Barnich:2009se, Cachazo:2014fwa}, and correlators of chiral operators in 4d and 6d superconformal field theories \cite{Beem:2013sza, Beem:2014kka}. 

There is a small handful of Virasoro blocks that has been computed exactly. Minimal model blocks obey ordinary differential equations, and in the simplest cases are just hypergeometric functions \cite{bpz}. Correlators of degenerate fields in non-minimal models, such as Liouville theory, also obey ODEs. At $c=1$ and $c=25$, Zamolodchikov derived the blocks for generic internal dimension but specific external dimensions \cite{Zamolodchikov:426555}. More recently, all $c=1$ blocks have been understood as Fourier coefficients of a series expansion of the Painlev\'e tau function \cite{Gamayun:2012ma}. There are also some exact results at large $c$  \cite{fkw1, fkw2}, to be discussed below. 

This distinguished set aside, one must resort to some form of perturbative expansion of the blocks. In \cite{Alba:2010qc}, a formula was derived for the coefficient of an arbitrary term in a series expansion in the conformal cross-ratio. This is a combinatoric formula, inspired by the AGT correspondence \cite{Alday:2009aq} with the instanton partition function of a 4d $\N=2$ gauge theory. In this work, we will not only derive these series coefficients from a purely 2d perspective, but will provide three closed-form expansions of the Virasoro blocks that go {\it beyond} the series expansion, for generic values of external and internal operator dimensions and the central charge. These should be viewed as partial resummations of the series, which utilize the rich substructure of Virasoro representations. 

There are traditionally two methods for computing the Virasoro blocks. The first is to construct the blocks level-by-level using the Virasoro algebra, which quickly becomes tiresome and reveals few general patterns. The second is to use two recursion relations derived by Zamolodchikov \cite{zamo, zamo2}. His main insight was that the generic blocks are strongly constrained by the structure of degenerate representations of the Virasoro algebra. This leads to two nested recursive structures for the blocks, built to generate the proper poles in either the central charge or the conformal dimension of the internal operator. These recursion relations are widely used to generate a series expansion in either the conformal cross-ratio, or the uniformizing coordinate of the Riemann sphere with singular points deleted, respectively.

In what follows, we will solve both of these recursion relations for generic parameters, yielding a pair of closed-form representations of the Virasoro blocks. The first of these representations is particularly intuitive: it is the decomposition of the Virasoro block into the hypergeometric global conformal blocks, defined with respect to the $SL(2,\mathbb{R})$ chiral half of the global conformal symmetry, $SO(2,2) \simeq SL(2,\mathbb{R}) \times SL(2,\mathbb{R})$. We give the coefficients in this decomposition at arbitrary level above the primary state. These have a clean physical interpretation as sums of squared OPE coefficients of quasi-primary operators. The second representation\footnote{After the present work was released, we learned that this representation was also given in \cite{Ribault:2014hia}. We thank Sylvain Ribault for bringing this to our attention.} is in terms of the uniformizing coordinate mentioned above, which covers the entire $z$ plane away from the singularities of the correlator. Consequently, one may wish to view this representation as our most complete representation of the Virasoro block. 

We also provide a third representation of the generic Virasoro block, by way of a semiclassical limit. To explain how this works, we recall the well-known fact that the global block is also the large $c$ limit of the Virasoro block with all dimensions held fixed. Therefore, one may view the global decomposition of the Virasoro block for generic $c$ as a sum over large $c$ blocks, with coefficients that re-sum all $1/c$ corrections to the limit. This perspective generalizes nicely to other semiclassical limits, in which some of the internal and external dimensions scale with large $c$: assuming one knows the Virasoro block in such limits, the {\it generic} block can again be written as a sum over the limit blocks, with coefficients that restore finite $c$. In the ``heavy-light'' semiclassical limit studied recently in \cite{fkw2}, in which one pair of external operator dimensions scales linearly with large $c$, the blocks are known. We thus give a third representation of the Virasoro block, which we refer to as the ``semiclassical representation'', as a sum over these semiclassical blocks with closed-form coefficients at arbitrary level. 

We note that all of these results also double as torus one-point Virasoro blocks, which are special cases of sphere four-point Virasoro blocks. 

It is straightforward to expand the Virasoro blocks in $1/c$. The expansion of the global representation contains information about loop corrections to scattering of four light particles in AdS$_3$ quantum gravity, where the coupling constant is Newton's constant, $G_N = 3R_{\rm AdS}/2c$. The expansion of the semiclassical representation, on the other hand, contains corrections to the emergent thermality of heavy microstates at large $c$ \cite{fkw2}. The details of these connections are sure to be interesting. We leave their full investigation for future work, for now giving some general discussion in Section \ref{v}. We also apply some recent results on the large $c$ vacuum block to the computation of R\'enyi entropy in 2d CFT for generic interval size. 

The paper is organized as follows. In Section \ref{ii}, we recall salient aspects of 2d CFTs; define the Virasoro blocks; and review, then solve, Zamolodchikov's two recursion relations. In Section \ref{iii}, we give a third representation of the blocks via the heavy-light semiclassical limit. Section \ref{v} discusses the relation of our and other results on Virasoro blocks to particle scattering and thermal physics in AdS$_3$/CFT$_2$ and to computation of entanglement in large $c$ CFTs; we also outline directions for future work. In Appendix \ref{AppB} we prove that many of the residues entering into the Zamolodchikov recursion formulae actually vanish when the external operator dimensions are pairwise identical, and that even more residues vanish for the particular case of the vacuum block. The remaining appendices contain details that may be used to verify some formulae in the text.

\sec{Conformal blocks in two dimensions}\label{ii}
All two-dimensional conformal field theories contain a stress tensor that generates two copies of the Virasoro algebra \cite{bpz, Ginsparg:1988ui, DiFrancesco:1997nk}. Focusing on the holomorphic sector in complex coordinates $(z,\zb)$, the holomorphic stress tensor $T(z) = T_{zz}$ has a mode expansion $T(z) =  \sum_{n\in\mathbb{Z}}{L_n z^{-n-2}}$, where the modes obey
\e{cbb}{[L_m,L_n] = (m-n)L_{m+n} + {c\over 12}n(n^2-1)\delta_{m+n,0}~.}
A given CFT has a spectrum of local primary operators labeled by left- and right-moving conformal dimensions $(h,\hb)$; these are the eigenvalues under $(L_0,\overline{L}_0)$, respectively. A primary $\Oc$ is annihilated by all positive modes of both Virasoro algebras. The action of the negative modes generates the highest-weight representation spanned by all Virasoro descendants of $\Oc$, also known as the Verma module, which we denote $M(c,h)\otimes M(\overline{c},\hb)$.\footnote{In the presence of null states, $M(c,h)$ stands for the irreducible module.} The Hilbert space of the CFT can be written as the direct sum
\e{cbc}{{\cal H} = \bigoplus_{h,\hb}M(c,h) \otimes M(\overline{c},\hb)~.}

The operator product expansion (OPE) of two primary fields $\Oc_1$ and $\Oc_2$ is
\e{ope}{\Oc_1(z,\zb)\Oc_2(0,0) = \sum_pC^p_{~12}|z|^{2(h_p-h_1-h_2)}\Psi_p(z,\zb|0,0)~.}
The sum is over all local primaries in the CFT, with OPE coefficients $C^p_{~12}$, and $\Psi_p$ includes all contributions of the $p$'th conformal family, $M(c,h_p)$. $\Psi_p$ is often written as \cite{bpz}
\e{psi}{\Psi_p(z,\zb|0,0) = \Big|\sum_{\lbrace k \rbrace} \b_{12}^{p\lbrace k \rbrace} z^K\Oc_p^{\lbrace k \rbrace}(0)\Big|^2}
where the sum is over all descendants $\Oc_p^{\lbrace k \rbrace}$ at level $K=\sum k$, with $\lbrace k \rbrace$ indexing the $p(K)$ descendants. The $\b$'s are coefficients; they are not known in closed form, but are nevertheless fully determined by Virasoro symmetry. 

Consider a four-point function of primaries, $\langle\Oc_1(z_1)\Oc_2(z_2)\Oc_3(z_3)\Oc_4(z_4)\rangle$, suppressing dependence on $\zb$ for simplicity. Using global conformal symmetry to put operators at $z_1=\infty, z_2=1, z_4=0$ and assigning $z_3= z$, the conformally-invariant cross ratio $x$ is
\e{cbd}{x={z_{12}z_{34}\over z_{13}z_{24}} = z}
where $z_{ab}:= z_a-z_b$. Expanding in the $x\rar 0$ OPE channel, the four-point function is constrained to take the form
\es{cbe}{\langle\Oc_1(\infty)\Oc_2(1)\Oc_3(z)\Oc_4(0)\rangle &= \lim_{w\rar\infty}|w|^{4h}\langle\Oc_1(w)\Oc_2(1)\Oc_3(z)\Oc_4(0)\rangle\\
&= {1\over |z|^{2(h_3+h_4)}}G(z,\zb)}
where
\e{cbf}{G(z,\zb)=\sum_p C_{12p}C^p_{~34} |\cF(c,h_i,h_p,z)|^2}
We are using a shorthand $h_i:= \lbrace h_1,h_2,h_3,h_4\rbrace$, and the index is raised by the Zamolodchikov metric, $g_{ab}= \langle\Oc_a|\Oc_b\rangle $.

The object $\cF(c,h_i,h_p,z)$ is the sought-after Virasoro conformal block. It contains all contributions to the four-point function from Virasoro descendants of the primary $\Oc_p$. 
The block admits a series expansion around $z=0$,
\e{cbh}{\cF(c,h_i,h_p,z) = z^{h_p} \sum_{K=0}^{\infty} \cF_K(c,h_i,h_p) z^K~.}
The $\cF_K$ may be computed level-by-level using the OPE \eqr{ope}. To do so requires explicit knowledge of $\Psi_p$, and hence the coefficients $\b$. 

We now invoke the $SL(2,\mathbb{R})\times SL(2,\mathbb{R})$ global conformal subalgebra, generated on the left by the modes $\lbrace L_{\pm 1}, L_0\rbrace$ and likewise on the right. All modules $M(c,h_p)$ contain an infinite number of quasi-primary operators, which are primary only with respect to the global conformal symmetry. In the OPE, one is free to group terms by global conformal families, generated by powers of $L_{-1}$ acting on a quasi-primary.\footnote{Indeed, this was done in the original treatment in Appendix B of \cite{bpz}.} This is helpful because global descendant contributions to the OPE {\it are} known \cite{Bowcock:1990ku}. This $SL(2,\mathbb{R})$ decomposition, in turn, induces a branching of the Virasoro block into global blocks, which are just hypergeometric functions \cite{Ferrara:1974ny}:
\e{qp}{\cF(c,h_i,h_p,z) = \sum_{q} {C'_{12q}C'_{34q}\over \N_q} \,z^{h_q}\hyp(h_q+h_{12},h_q+h_{34}, 2h_q;z)~.}
This sum is over all quasi-primaries $O_q$ in the Verma module $M(c,h_p)$. We have explicitly written the norms $\N_q=\langle O_q|O_q\rangle$ to emphasize that the Zamolodchikov metric for quasi-primaries is nontrivial. The expansion coefficients $C'_{12q}$ and $C'_{34q}$ are simply the OPE coefficients of $O_q$, normalized by those of $\Oc_p$:
\e{cbia}{C'_{12q} = C_{12q}/C_{12p}~.}
Deriving this from \eqr{psi} is somewhat tedious. A more intuitive method is to consider a four-point function of chiral operators, where the anti-chiral block trivializes. The four-point function  can be expanded in global blocks instead of local blocks; grouping the former by conformal family leads directly to \eqr{qp}--\eqr{cbia}. Because $\cF$ is chiral, this holds in general.

For what follows, it will be very useful to organize the quasi-primary sum \eqr{qp} by level. A level-$q$ quasi-primary in the Verma module $M(c,h_p)$ has dimension $h_q=h_p+q$, so we can write the global decomposition \eqr{qp} as
\e{qp2}{\cF(c,h_i,h_p,z) = z^{h_p} \sum_{q=0}^{\infty} \chi_q(c,h_i,h_p)\,z^q \hyp(h_p+q+h_{12},h_p+q+h_{34}, 2(h_p+q);z)~,}
where the expansion coefficient $\chi_q$ is 
\e{chiq}{\chi_q(c,h_i,h_p) = \sum_{O_q\, \in\, M(c,h_p)}{C'_{12q}C'_{34q}\over \N_{q}}~.}
That is, $\chi_q$ is just the sum over squared OPE coefficients of all level-$q$ quasi-primary descendants of $\Oc_p$, normalized as indicated. Again, these are fixed by Virasoro symmetry, but neither the OPE coefficients nor the norms are known at arbitrary level $q$. All blocks have $\chi_0=1$ and $\chi_1=0$. 

\vs

It will be useful to have an example at hand. Let us consider the vacuum Virasoro block,
\e{vblock}{\cF_{\rm vac}(c,h_i,z) := \lim_{h_p\rar 0}\cF(c,h_i,h_p,z)~.}
Fusion onto the identity requires a pairwise correlator, $h_{12}=h_{34}=0$, so
\e{cbj}{\cF_{\rm vac}(c,h_i,z) = \sum_{q=0}^{\infty} \chi_{\vac,q}(c,h_i) \,z^q\hyp(q,q,2q;z)~.}
In this case, the OPE coefficients $C_{12p}$ and $C_{34p}$ are norms. It is straightforward to enumerate the quasi-primaries and compute their norms and OPE coefficients, which we do in Appendix \ref{AppA}. The nonzero terms through $q=7$ are
\bea\label{cvac}
&&\!\!\!\!\!\!\chi_{\vac,2} = {2h_1h_3\over c}\nonumber\\
&&\!\!\!\!\!\!\chi_{\vac,4}= {10\left(h_1^2+{h_1\over 5}\right)\left(h_3^2+{h_3\over 5}\right)\over c(5c+22)}\\
&&\!\!\!\!\!\!\chi_{\vac,6}= \frac{(14 h_1^2+h_1) (14 h_3^2+h_3)}{63 c (70 c+29)}\nonumber\\&&+ \frac{4 h_1 h_3 \left(c(70  h_1^2+42  h_1+8) +29 h_1^2-57 h_1-2\right) \left(c(70 h_3^2+42  h_3+8) +29 h_3^2-57 h_3-2\right)}{3 c (2 c-1) (5 c+22) (7 c+68) (70 c+29)\nonumber}
\eea

Notice that these diverge at minimal model values of $c$ at which the exchanged quasi-primary becomes null. The constraints of Virasoro symmetry when $c<1$ form the backbone of Zamolodchikov's recursion relations for the Virasoro blocks, to which we now turn.

\subsec{Virasoro blocks I: Global representation}\label{ii-i}
In a pair of papers \cite{zamo, zamo2}, Zamolodchikov presented two recursion relations for the Virasoro conformal blocks.\footnote{Nice reviews can be found in e.g. \cite{Zamolodchikov:1990ww, Runkel:2001ng}; we warn the reader of notational and conventional differences.} Zamolodchikov's main observation was that, viewed as analytic functions of $c$ or $h_p$, the blocks have poles, due to the existence of null states in the space of Virasoro highest weight representations. The sum over these poles, multiplied by the correct residues, generates an algorithm for computing the Virasoro block in general. We begin with the recursion with respect to poles in the $c$-plane, giving a detailed explanation of the origins of its components along the way. 

The $c$-recursion relation is as follows: 
\es{rec}{\cF(c,h_i,h_p,z) &= z^{h_p}\hyp(h_p+h_{12}, h_p+h_{34}, 2h_p; z) \\&+ \sum_{m\geq 1, n\geq 2}^{\infty} {R_{mn}(h_i,h_p)\over c-c_{mn}(h_p)} \cF(c_{mn}(h_p), h_i, h_p+mn, z)~.}
This converges in the unit disk, $|z|<1$. The ``seed'' of the recursion is the Virasoro block at $c\rar\infty$ with all dimensions held fixed:
\e{sc1}{\lim_{c\rar\infty} \cF(c,h_i,h_p,z) = z^{h_p}\hyp(h_p+h_{12},h_p+h_{34},2h_p;z)~.}
This is the well-known fact that the Virasoro block reduces to the global block in such a limit: the Virasoro algebra fixes states built from $k$ modes $L_{-n}$ with $n>1$ to have norms that scale as $c^k + O(c^{k-1})$, so their contribution to \eqr{qp2} vanishes at large $c$. \eqr{sc1} is exact in $z$ within the unit disk. 

In deriving \eqr{rec}, Zamolodchikov's starting point was that the Virasoro block has poles at minimal model values of $c$, where exchanged quasi-primary descendants of $\Oc_p$ become null. Specifically, the $c_{mn}(h_p)$ are the minimal model values of $c$ where a level $mn$ descendant becomes null. These are conveniently parameterized as 
\e{rca}{c_{mn}(h_p) = 13-6\left(t_{mn}(h_p) + {1\over t_{mn}(h_p)}\right)\leq 1}
where
\e{rcb}{t_{mn}(h_p) = {2h_p+mn-1+\sqrt{4h_p(h_p+mn-1)+(m-n)^2}\over n^2-1}>0~.}
These define the zeroes of the Kac determinant, where $n$ and $m$ are conventionally labeled $r$ and $s$, respectively, in the classification of minimal model representations $\phi_{(r,s)}$. In the definition of $t_{mn}$, we restrict to the positive branch of the square root. 

The numerators in \eqr{rec} are to be thought of as residues at $c=c_{mn}(h_p)$. There are two parts to understanding their form. First, because the level $mn$ null state is the primary state of its own Verma submodule, the residue of its pole is proportional to the Virasoro block for the submodule itself; this explains the appearance of  $\cF(c_{mn}(h_p), h_i, h_p+mn, z)$ in the residue. Furthermore, the residue must vanish when the central charge equals $c_{mn}(h_p)$, because the null state actually resides in the spectrum of the theory. Accordingly, the residues' dependence on external dimensions $h_i$ must be such that the residues vanish when the $h_i$ belong to the Kac table of minimal model dimensions. This structure is nicely exhibited in $\chi_{\vac,4}$ in \eqr{cvac}: the pole sits at $c=-22/5=c_{14}(0)$, and the residue vanishes when $h_1$ and $h_3$ equal 0 or $-1/5$. These are precisely the central charge and operator spectrum of the Yang-Lee minimal model, ${\cal M}(5,2)$. 

Using these considerations, the precise definition of the $R_{mn}$ requires slightly more patience, and some degree of faith. $R_{mn}$ factorizes as
\e{rcc}{R_{mn}(h_i,h_p) = A_{mn}(h_p) P_{mn}(h_i,h_p)~.}
Define
\es{rcd}{\ell_{jk}(m,n,h_p) &= {j-k \cdot t_{mn}\over \sqrt{t_{mn}}}\\
\ell_i(m,n,h_i,h_p) &= \sqrt{h_i+{\ell^2_{11}\over 4}}~.}
The full dependence in the external dimensions $h_i$ lies in the $\ell_i$. Then $P_{mn}$ and $A_{mn}$ were found to be
\es{pmn}{P_{mn}(h_i,h_p) &= \prod_{j,k}\left(\ell_2+\ell_1-{\ell_{jk}\over 2}\right) \left(\ell_2-\ell_1-{\ell_{jk}\over 2}\right)\left(\ell_3+\ell_4-{\ell_{jk}\over 2}\right) \left(\ell_3-\ell_4-{\ell_{jk}\over 2}\right)\\
A_{mn}(h_p) &= -\half \left({24 \left(t_{mn}^{-1}- t_{mn}\right)\over (m^2-1) t_{mn}^{-1} - (n^2-1) t_{mn}}\right) \prod_{a,b}{1\over \ell_{ab}}~.}
The products run over the following ranges,
\es{range}{j&=-m+1,-m+3,\ldots, m-3,m-1\\
k&=-n+1,-n+3,\ldots, n-3,n-1\\
a&=-m+1,-m+2,\ldots, m-1,m\\
b&=-n+1,-n+2,\ldots, n-1,n}
with the exception of the pairs $(a,b) = (0,0)$ and $(m,n)$. The factor in parentheses in $A_{mn}$ equals the Jacobian $d h_p/d c_{mn}(h_p)$. 

The structure of $P_{mn}$ is precisely what is needed to remove all null states from the minimal model Virasoro blocks. (Note that $R_{mn}$ should be polynomial in $h_i$ because it must reproduce result of the OPE, which only ever produces non-negative integer powers of $h_i$; it is unobvious from its definition in \eqr{pmn}, but $P_{mn}$ does indeed satisfy this requirement.\footnote{To see this, consider the product over the factors $(\ell_2+\ell_1-{\ell_{jk}\over 2})(\ell_2-\ell_1-{\ell_{jk}\over 2}) = (\ell_2-{\ell_{jk}\over 2})^2- \ell_1^2$. Because $\ell_{-j,-k}=-\ell_{jk}$ and the ranges for $(j,k)$ are symmetric about $j=k=0$, the product over $(j,k)$ only produces even powers of each $\ell_i$, and hence integer powers of each $h_i$. For unequal $h_i$, the polynomial is of degree $2mn$; for each pair of equal $h_i$, the degree decreases by $mn/2$.}) On the other hand, $A_{mn}$ has never been derived from first principles to our knowledge, but has passed extensive checks against series expansions of the block and Liouville computations \cite{Zamolodchikov:2003yb}.

To further compactify notation, we introduce one more definition:
\es{gam2}{\gamma_{mn}(c,h_i,h_p) := {R_{mn}(h_i,h_p)\over c-c_{mn}(h_p)}~,}
so that the recursion is simply
\es{rec2}{\cF(c,h_i,h_p,z) &= z^{h_p}\hyp(h_p+h_{12}, h_p+h_{34}, 2h_p; z) \\&+ \sum_{m\geq 1, n\geq 2}^{\infty}\gmn(c,h_i,h_p) \cF(c_{mn}(h_p), h_i, h_p+mn, z)~.}

Before solving this recursion relation, let us make a few remarks about it.

When even a single pair of $h_i$ is equal, $P_{mn}=0$ for $mn$ odd. This is because $\ell_{00}=0$ appears in the product defining $P_{mn}$. For generic $h_p$, this cannot be cancelled by a pole in $A_{mn}$, so
\e{resv}{R_{mn}(h_i,h_p)=0~~\text{for}~mn~\text{odd , } ~ h_1=h_2~.}
Subsequently, the recursion \eqr{rec2} implies that the Virasoro block $\cF$ receives no contributions at odd levels above the primary. This is consistent with the fact that, for example, an odd-spin quasi-primary cannot appear in the OPE of two identical scalar quasi-primaries.   

Additionally, many residues vanish when computing $\cF_{\vac}$. For pairwise identical operator dimensions, $h_1=h_2, h_3=h_4$, the following three properties hold:
\begin{subequations}
\bea\label{three}
\text{\bf i)}&&\!\!\!\!R_{mn}(h_i,0)=0~~\text{for}~mn~\text{odd}\\
\text{\bf ii)}& &\!\!\!\!R_{mn}(h_i,0)=0~~\text{for}~ m\geq n\label{second}\\
\text{\bf iii)}&&\!\!\!\!R_{mn}(h_i,0) = 0~~\text{for}~h_1=0\,,~\text{all}~(m,n)~.\label{third}
\eea
\end{subequations}
This is shown in Appendix \ref{AppB}. 

Finally, we emphasize that the recursion is linear in the hypergeometric function. This implies that iteration of the recursion yields a linear combination of the global blocks. In particular, the recursion's output is precisely the coefficients $\chi_q$ in the global decomposition of the full Virasoro block \eqr{qp2}.
\subsubsection{Solution}\label{ii-i-i}

We now present the Virasoro block in a closed-form expansion. It takes the form of the global decomposition \eqr{qp2}. The coefficients $\chi_q$ for $q\geq 2$ are order-$\lfloor {q\over 2}\rfloor$ polynomials in the $\g_{mn}$ coefficients defined in \eqr{gam2}:
\e{sol1}{\chi_q(c,h_i,h_p) = \sum_{j=1}^{\lfloor {q\over 2}\rfloor}\prod_{\ell=1}^j\sum_{\substack{m_{\ell}\,\geq \,1,\\n_{\ell}\,\geq \,2}}^{\infty} \g_{m_{\ell}n_{\ell}}\big(c_{\rm eff}^{(\ell)},h_i,h_{p,\,{\rm eff}}^{(\ell)})\,\Big|_{\sum\limits_{\ell=1}^j m_\ell n_\ell=q}~.}
We have introduced an effective central charge $c_{\rm eff}^{(\ell)}$ and internal operator dimension $h_{p,\,{\rm eff}}^{(\ell)}$. To define these, let us first define the following ``anomalous dimension'',
\e{rcf}{\D h_p^{(\ell)}  := \sum_{r=1}^{\ell}m_{r}n_{r}~.}
Then 
\e{hpeff}{h_{p,\,{\rm eff}}^{(\ell)}:= h_p+\D h_p^{(\ell-1)}}
and
\e{ceff}{c_{\rm eff}^{(\ell)} := c_{m_{\ell-1}n_{\ell-1}}(h_{p,\,{\rm eff}}^{(\ell-1)})~.}
We have assigned $c_{\rm eff}^{(1)}:= c$.

Using \eqr{sol1} and the series representation of the hypergeometric function, it is straightforward to write down a closed-form expression for an arbitrary coefficient in the $z$-expansion \eqr{cbh}:
\e{cbi}{\cF_K(c, h_i, h_p) = \sum_{q=0}^K \chi_q (c, h_i, h_p) {(h_p+q+h_{12})_{K-q}(h_p+q+h_{34})_{K-q}\over (K-q)! (2h_p+2q)_{K-q}}}
where $(a)_n$ is the rising Pochhammer symbol, and $\chi_q$ are defined in \eqr{sol1}. Of course, the representation \eqr{qp2} is more efficient, because it re-sums the effects of the global descendants. 
\vs

Let us unpack this formula a bit. \eqr{sol1} is a sum, indexed by $j$, of $\lfloor {q\over 2}\rfloor$ terms. Each of these is a $j$-fold nested sum over a product of $j$ coefficients $\g_{m_{\ell}n_{\ell}}$ with $1\leq \ell \leq j$, whose arguments are themselves functions of the indices $(m_r,n_r)$ with $r<\ell$. The products are subject to the constraint that they contribute to $\chi_q$, which fixes $\sum_r m_rn_r=q$. 

The relation of \eqr{sol1} to the recursion \eqr{rec2} is manifest. In \eqr{sol1}, $\ell$ counts the level of the recursion from which $\g_{m_{\ell}n_{\ell}}$ comes. There are $\lfloor {q\over 2}\rfloor$ terms in the sum because the power of $z$ always jumps by at least two ($m_{\ell}n_{\ell}\geq 2$ for all $\ell$). For each iteration of \eqr{rec2}, the effective dimension of the internal operator at level $\ell$ increases by $m_{\ell-1}n_{\ell-1}$. This motivates our definition of $\D h_p^{(\ell)}$ as an anomalous dimension at level $\ell$: in particular, the effective dimension of the internal operator at level $\ell$ is shifted by this amount. The effective central charge at each level of recursion is the degenerate central charge \eqr{rca} defined at the previous level, which explains our definition of $c_{\rm eff}^{(\ell)}$. 

To gain some intuition, it is instructive to write out \eqr{sol1} at low levels, which we do in Appendix \ref{appc}, and compare to the expansion of the recursion \eqr{rec2}. For now, some comments are in order:%

\vs
\bul Independent of its role in defining $\cF$, \eqr{sol1} gives the sum over level-$q$ quasi-primary OPE coefficients in \eqr{chiq}. This is especially interesting because the set of quasi-primaries -- that is, the orthogonal basis at arbitrary level $q$ -- is not known, much less their OPE coefficients and norms. It would be very interesting to derive a generating functional for these quasi-primaries. This would amount to diagonalization of the Gram matrix at arbitrary level.  A more specific goal would be to find this generating functional for the vacuum block, perhaps in a $1/c$ expansion. This has a holographic interpretation as a generating functional for the orthogonal basis of graviton states in AdS$_3$ quantum gravity.  

\bul Because $c$ only appears in \eqr{sol1} at $\ell=1$, it is straightforward to expand $\cF$ around some fixed $c=c_*$: simply Taylor expand the definition of $\g_{m_1n_1}$ in \eqr{gam2}. For example, we can immediately write down the full set of $1/c$ corrections to the $c\rar\infty$ limit of $\cF$ with all dimensions held fixed. Expanding $\cF$ around $c\rar\infty$ as
\es{cm1}{\cF(c\rar\infty,h_i,h_p,z) &= z^{h_p} \hyp(h_p+h_{12},h_p+h_{34}, 2h_p;z)+ \sum_{n=1}^{\infty}c^{-n}\cF^{(n)}(h_i,h_p,z)}
the $O(c^{-n})$ term is
\e{cm1a}{\cF^{(n)}(h_i,h_p,z) = z^{h_p}\sum_{q=2}^{\infty}\chi_q^{(n)}(h_i,h_p)\, z^q\hyp(h_p+q+h_{12},h_p+q+h_{34}, 2(h_p+q);z)}
where $\chi_q^{(n)}(h_i,h_p)$, the $O(c^{-n})$ part of $\chi_q$, is given again by \eqr{sol1} with the lone substitution
\e{cm2}{\g_{m_1n_1}(c,h_i,h_p)  ~\to~ {R_{m_1n_1}(h_i,h_p)\over c} \left({c_{m_1n_1}(h_p)\over c}\right)^{n-1}~.}
In Section \ref{v}, we will return to these corrections in the context of the semiclassical expansion of bulk scattering in AdS$_3$ quantum gravity. 

\bul The representation \eqr{sol1} obscures certain properties of the conformal block. As an example, consider the vacuum block. In the large $c$ expansion, it would naively seem that $\chi^{(1)}_{\vac, q}$ is nonzero for all $q$, because all $\g_{mn}$ contain a $1/c$ piece when expanded at large $c$. However, this is incorrect: only $\chi^{(1)}_{\vac,2}$ is nonzero, and
\e{cm3}{\cF_{\vac}(c\rar\infty,h_i,z) \approx 1 + {2h_1h_3 \over c} \,z^2 \hyp(2,2,4;z) + O\left({1\over c^2}\right)~.}
From the perspective of the global decomposition, this is obvious. A quasi-primary made of $p$ stress tensors contributes to $\cF_{\vac}$ at $O(c^{-p})$ and beyond; the only quasi-primary made of a single stress tensor is $T$ itself, and the $1/c$ term in \eqr{cm3} corresponds to $T$ exchange. But from the formula \eqr{cm1a}, this requires non-trivial cancellations among the $1/c$ pieces of each term at fixed $q>2$.

\subsection{Virasoro blocks II}\label{ii-ii}
Our solution can also be applied to other recursion relations of the same underlying structure. We briefly highlight two other recursion relations for Virasoro conformal blocks. 

\subsubsection{Elliptic representation}\label{ii-ii-i}
The first is the recursion of \cite{zamo2} which sums over poles in $h_p$ rather than poles in $c$. The explicit representation of the block is\footnote{Note that our convention \eqr{cbe} differs from \cite{zamo2} by the overall power of $z$.}
\e{h1}{\cF(c,h_i,h_p,z) = (16q)^{h_p-{c-1\over24}}z^{{c-1\over24}}(1-z)^{{c-1\over24} -h_2-h_3}\theta_3(q)^{{c-1\over2}-4\sum_i h_i}H(c,h_i,h_p,q)~.}
The elliptic variable $q$ maps the four-punctured sphere to the upper-half plane by\footnote{Inversely, $z$ is the modular lambda function: $z = \lambda(\tau)$.} 
\e{h2}{q = e^{i\pi  \tau}~, \quad \tau = {i} {K(1-z)\over K(z)}~.}
$K(z)$ is the complete elliptic integral of the first kind, which has a hypergeometric representation, $K(z) = {\pi\over 2} \hyp(\half,\half,1;z)$. $\theta_3(q)$ is the Jacobi theta function, 
\e{h3}{\theta_3(q) = \sum_{n=-\infty}^{\infty}q^{n^2}~.}
The function $H(c,h_i,h_p,q)$ admits a  power series expansion around $q=0$, 
\e{sol2}{H(c,h_i,h_p,q) = \sum_{N=0}^{\infty} H_N(c,h_i,h_p) q^N}
where the $H_N$ are obtained from a recursion formula,
\e{Hrec2}{H(c,h_i,h_p,q) = 1+\sum_{\substack{m\,\geq \,1,\\ n\,\geq\, 1}}^{\infty}{(16q)^{mn}\hat R_{mn}(c,h_i)\over h_p-h_{p,mn}(c)}H(c,h_i,h_{p,mn}+mn,q)~.}
The residues $\hat R_{mn}$ are equal to the $R_{mn}$ defined in \eqr{rcc} upon removing the Jacobian factor in parenthesis in \eqr{pmn}. The $h_{p,mn}(c)$ are the dimensions of the degenerate representations at level $mn$, defined by solving the relations \eqr{rca}, \eqr{rcb} for $h_p$ as a function of $c$; explicitly,
\e{}{h_{p,mn}(c) = {1\over 4} (n^2-1) t(c) + {1\over 4} (m^2-1){1\over t(c)} - \half(mn-1)}
where
\e{}{t(c)= 1+{1\over 12}\left(1-c\pm\sqrt{(1-c)(25-c)}\right)}

Let's again introduce a notation for the coefficient of $q^{mn}$ as
\e{erba}{\zeta_{mn}(c,h_i,h_p)  := {16^{mn}\hat R_{mn}(c,h_i)\over h_p-h_{p,mn}(c)}}
so that
\e{Hrec}{H(c,h_i,h_p,q) = 1+\sum_{\substack{m\,\geq \,1,\\ n\,\geq\, 1}}^{\infty}q^{mn}\zeta_{mn}(c,h_i,h_p)H(c,h_i,h_{p,mn}+mn,q)~.}

This recursion is simpler than the global recursion \eqr{rec}. With each iteration, only the $h_p$ dependence changes, as opposed to both $h_p$ and $c$. And because $h_{p,mn}$ appears on the right-hand side of \eqr{Hrec} rather than some function of $h_{p}$, the effective internal dimension at level $\ell$ of recursion depends only on level $\ell-1$, as opposed to its value at all previous levels; contrast this with \eqr{hpeff}. Consequently, taking $h_p = h_{p,mn}+mn$ turns \eqr{Hrec} into an infinite-dimensional matrix equation for the elements $H(c,h_i,h_{p,mn}+mn,q)$, of the form
\e{mateq}{ \vec{H} = \vec{1} + {\bf M}\cdot \vec{H}~.}
$\vec{H}$ is a vector of the $H(c,h_i,h_{p,mn}+mn,q)$, and ${\bf M}$ is linear in the $\zeta_{mn}$. Feeding its solution back into \eqr{Hrec} yields $H(c,h_i,h_p,q)$ for generic $h_p$. In practice, the matrix equation is solved with a cutoff at some level $L$, thus yielding the block in a $q$-expansion up to $O(q^L)$.\footnote{By expanding to $O(q^4)$, it is easy to prove that there are two, and only two, cases where ${\bf M}$ vanishes identically for generic internal dimensions. These are the cases studied in \cite{Zamolodchikov:426555}: $c=1$ for identical external operators with $h=1/16$, and $c=25$ for identical external operators with $h=15/16$.}

\vs
We can now present the expression for an arbitrary expansion coefficient $H_N$:
\e{sol2a}{H_N(c,h_i,h_p) = \sum_{j=1}^N\prod_{\ell=1}^j\sum_{\substack{m_{\ell}\,\geq \,1,\\ n_{\ell}\,\geq\, 1}}^{\infty}\zeta_{m_{\ell}n_{\ell}}(c, h_i, \hat h_{p,{\rm eff}}^{(\ell)})\Big|_{\sum_{\ell=1}^j m_{\ell}n_{\ell}= N}~.}
The effective internal dimension, now denoted $\hat h_{p,\,{\rm eff}}^{(\ell)}$, is
\e{hpeffH}{\hat h_{p,\,{\rm eff}}^{(\ell)}:= h_{p,m_{\ell-1}n_{\ell-1}}+m_{\ell-1}n_{\ell-1}~.}
We assign $h_{p,m_0n_0}= h_p$ and $m_0=n_0=0$, so that $\hat h_{p,\,{\rm eff}}^{(1)}=h_p$. Taking $h_p=h_{p,mn}+mn$ for any $(m,n)$ gives the solution of the matrix equation \eqr{mateq}. 

Equations \eqr{h1}, \eqr{sol2} and \eqr{sol2a} provide a representation of $\cF$ which is convergent everywhere on the $z$ plane with the points $z=1,\infty$ excised \cite{zamo2}: the parameter $q$ is bounded as $|q|\leq 1$, with saturation only at $z=1,\infty$. Within $|z|<1$ the rate of convergence in $q$ is faster than \eqr{cbh}: at small $z$ for example, $q\approx z/16+O(z^2)$. 

\subsubsection{One-point block on the torus}\label{ii-ii-ii}

We can also write down a closed-form expansion of the torus one-point conformal block. On a torus of modular parameter $\tau={1\over 2\pi i} \log q$, the one-point function of a primary $\Oc$ is
\e{tor1}{\langle \Oc \rangle_{\tau} = \text{Tr}(q^{L_0-c/24}\overline{q}^{\overline{L}_0-\overline{c}/24}\Oc)~.}
The trace is over the full CFT Hilbert space. This can be written as
\e{tor3}{\langle \Oc \rangle_{\tau} =  \sum_{p} {\langle \Oc_p|\Oc|\Oc_p\rangle\over \N_p} |\cF_{\tau}(c, h, h_p, q)|^2}
where $\N_p = \langle \Oc_p|\Oc_p\rangle$. The sum runs over all primaries, and we have defined $\cF_{\tau}(c, h, h_p, q)$ as the holomorphic one-point Virasoro conformal block on the torus, with $q$-expansion normalized as
\e{tor4}{\cF_{\tau}(c, h, h_p, q) = q^{h_p-c/24} \sum_{K=0}^{\infty} \cF_{\tau,K}(c, h, h_p) q^K}
where $\cF_{\tau,0}(c, h, h_p) =1$. When $\Oc={\bf 1}$, \eqr{tor4} is the Virasoro character of the Verma module $M(c,h_p)$, and \eqr{tor3} reduces to the torus partition function. 

It is known that $\cF_{\tau}$ is a special case of the sphere four-point block \cite{Fateev:2009me, Poghossian:2009mk, Hadasz:2009db}. Therefore, under such a mapping, we have also derived $\cF_{\tau}$ using either the global block expansion \eqr{sol1} or the elliptic expansion \eqr{sol2a}.\footnote{For instance, $\cF_{\tau}$ obeys a recursion relation that differs from the elliptic recursion for the sphere four-point block only by trivial factors in the definition of $H$; the exact expression can be found on pp.7-8 of \cite{Hadasz:2009db}. } 

\section{Virasoro blocks III: The heavy-light semiclassical limit}\label{iii}

In this Section we focus on a semiclassical limit motivated by AdS/CFT, dubbed the ``heavy-light'' semiclassical limit in \cite{fkw2}, in which the Virasoro block is known. This enables us to write another closed-form series expansion for the {\it generic} Virasoro block, and to study $1/c$ corrections to its semiclassical limit with ease.

The heavy-light limit is a large $c$ limit in which two external dimensions $(h_3,h_4)$ remain fixed (``light'' operators), and two dimensions $(h_1,h_2)$ scale linearly with $c$ (``heavy'' operators). Define
\es{scb}{h_{3} &:= h_L~, \quad h_{4} := h_{L}-\d_L\, ,\\
h_{1} &:= \eta c~, \quad h_{2} := h_1 -\d_H~.}
We then take the limit $c\rar\infty$ with $\eta$ held fixed, and $\d_L,\d_H \sim O(c^0)$. The remarkable result of \cite{fkw2} is that in this limit, the Virasoro block is essentially the global block written in a new coordinate $w(z)$, which is defined as 
\e{sca}{1-w= (1-z)^{\a}~, ~~ \text{where}~~\a= \sqrt{1-24 \eta}~.}
This coordinate defines a new background metric for the CFT generated by the heavy operators. Its invariance under an $e^{2\pi i /\alpha}$ rescaling of $1-z$ suggests an emergent thermality of heavy microstates at large $c$, and a holographic interpretation of heavy-light four-point correlators in terms of light particles propagating in conical defect or black hole geometries whose masses are fixed by $\a$ \cite{fkw2}. 

At leading order in this limit, the Virasoro block is, in our conventions, 
\e{scc}{\cFsc(h_i,h_p,w) = z^{2h_{L}-\d_L}(1-w)^{h_{L}(1-1/\a)}\left({w\over \a}\right)^{h_p-2h_{L}+\d_L}\hyp\big(h_p-{\d_H\over \a},h_p+\d_L,2h_p;w\big)~.}
where the superscript stands for ``semiclassical''. The first few terms in an expansion around $z=0$ are
\es{bla}{&\cFsc(h_i,h_p,w) \approx z^{h_p} \Bigg[1-{(\d_H-{h_p})(\d_L+{h_p})\over 2{h_p}}z\\
&+{1\over 4{h_p}(2{h_p}+1)}\Bigg((\d_L+{h_p})\Big(\d_H^2(1+\d_L+{h_p})-\d_H(1+\d_L+{h_p})(2{h_p}+1)\\&+{h_p}\Big(4\eta({h_p}-1)+({h_p}+1)^2+\d_L(1-12\eta+{h_p})\Big)\Big)+8\eta {h_p}(2{h_p}+1)h_{L}\Bigg)z^2+O(z^3)\Bigg]}

It was further observed in \cite{fkw2} that the Virasoro block $\cF$ can be obtained by a modified recursion formula in the mold of \eqr{rec2}: we now regard $\cF$ as an analytic function of $c$ with $h_L, \d_L, \eta, \d_H, h_p$ held fixed. Writing $\cF$ as a sum over its poles plus the piece at infinity yields a recursion formula, where the piece at infinity is $\cFsc$. It is important to emphasize that despite the appearance of $\cFsc$, this is a prescription for a {\it generic} Virasoro block that is simply well-suited to a $1/c$ expansion around the heavy-light semiclassical limit. 

The semiclassical recursion formula is then
\es{sce}{\cF(c,h_i,h_p,w)  &= \cFsc(h_i,h_p,w) + \sum_{m\geq 1, n\geq 2}^{\infty} \gmn^{\rm sc}(c,h_i,h_p) \cF(c_{mn}(h_p),h_i,h_p+mn,w)~.}
The expansion coefficients, which we label $\gsc$, differ from the $\gmn$ defined in the global recursion in \eqr{gam2} only by way of the relation $h_1=\eta c$. Namely, the residue at $c=c_{mn}(h_p)$ is now a function of $c$, leading to the following crucial substitutions:
\es{scf}{R_{mn}^{\sc}(h_i(c),h_p) &:= R_{mn}(h_i,h_p)\big|_{h_1 \rar \eta c_{mn}(h_p)}\\
\gsc(c,h_i,h_p) &:= \gmn(c,h_i,h_p)\big|_{h_1 \rar \eta c_{mn}(h_p)}~,}
where $h_2=h_1-\d_H$ is also a function of $c$ (cf. \eqr{scb}). 

It is now clear that the output of \eqr{sce} is a linear combination of semiclassical blocks,
\e{qp2sc}{\cF(c,h_i,h_p,w) = \sum_{q=0}^{\infty} \chi_q^{\rm sc}(c,h_i,h_p) \cFsc(h_i,h_p+q,w)}
where $\csc_0=1, \csc_1=0$. The coefficients $\csc_q$ can be read off immediately from our previous solution \eqr{sol1}:
\e{scg}{\csc_q(c,h_i,h_p) =\sum_{j=1}^{\lfloor {q\over 2}\rfloor}\prod_{\ell=1}^j\sum_{\substack{m_{\ell}\,\geq \,1,\\n_{\ell}\,\geq \,2}}^{\infty} \g^{\rm sc}_{m_{\ell}n_{\ell}}\big(c_{\rm eff}^{(\ell)},h_i,h_{p,\,{\rm eff}}^{(\ell)})\,\Big|_{\sum\limits_{\ell=1}^j m_\ell n_\ell=q}~.}
\eqr{qp2sc} and \eqr{scg} furnish yet another representation of the generic Virasoro block, which we label the ``semiclassical representation''. In Appendix \ref{appd}, we give the explicit expressions for $\chi_{\vac, q}^{\rm sc}$ for $q=2,4,6$, which may be used to check the veracity of the semiclassical representation against the known result \eqr{cvac}. Unlike the $\gmn$ in the global representation of the block, the $\gsc$ do not have a clean interpretation in terms of OPE coefficients in general, due to the dependence of $w$ on $\eta$.
\vs

The $1/c$ corrections to the heavy-light semiclassical blocks mimic the corrections to the 
blocks in the large $c$ limit with fixed dimensions, which we gave in \eqr{cm1}--\eqr{cm2}. The only difference here is that we now use the semiclassical expansion coefficients $\csc$. Expanding $\cF$ around the heavy-light semiclassical limit as
\e{sci}{\cF(c,h_i,h_p,z) \big|_{\rm sc~ limit} = \cFsc(h_i,h_p,w) + \sum_{n=1}^{\infty} c^{-n} \cF^{\sc,(n)}(h_i,h_p,w)}
and denoting the $O(c^{-n})$ part of $\csc_q$ as $\chi^{\sc, (n)}_q$, \eqr{qp2sc} yields
\e{scj}{\cF^{\sc,(n)}(h_i,h_p,w) = \sum_{q=2}^{\infty}\chi^{{\rm sc},(n)}_q(h_i,h_p)\cFsc(h_i,h_p+q,w)}
where $\chi_q^{\sc, (n)}(h_i,h_p)$ is given again by \eqr{scg} with the lone substitution
\e{sck}{\g^{\sc}_{m_1n_1}(c,h_i,h_p)  ~\to~ {R^{\sc}_{m_1n_1}(h_i(c),h_p)\over c} \left({c_{m_1n_1}(h_p)\over c}\right)^{n-1}~.}
We see that the perturbative corrections to $\cFsc(h_i,h_p,w)$ are weighted sums over other semiclassical blocks $\cFsc(h_i,h_p+q,w)$ for all possible $q\geq 2$. Note that they can also be written in terms of the ``thermal'' coordinate $w$.
\vs
In fact, in the case of the vacuum block, the semiclassical representation is more computationally efficient to implement than the more familiar representation \eqr{sol1}, as we now explain. This fact hinges on the property \eqr{third}, which states that the residues vanish whenever one pair of external dimensions vanishes. From \eqr{scf}, this implies that 
\e{sch}{c_{m_1n_1}(0)=0 \quad \Rightarrow \quad  \g_{m_1n_1}^{\rm sc}(c,h_i, 0)=0}
The key point is that at these values of $(m_1,n_1)$, the effective dimension of $\Oc_1$ appearing in $\g_{m_1n_1}^{\rm sc}$ is $h_1\rar \eta c_{m_1n_1}(0)=0$. These zeroes occur for the infinite set of pairs obeying $3m_1 = 2n_1-1$, where we have restricted to $m_1<n_1$ (cf. \eqr{second}).

The first such pair that is allowed is $(m_1,n_1)= (1,2)$. This is just the statement that the stress tensor never becomes null in any CFT. This zero alone eliminates slightly more than half of the terms in the sum \eqr{scg} when $h_p=0$; one can easily derive this using basic facts about compositions (i.e. ordered partitions), which we do in Appendix \ref{appd}. Removing the infinite tower of zeroes \eqr{sch} from the sum only improves efficiency further. 

Altogether, the representation \eqr{qp2sc}--\eqr{scg} gives a streamlined method for computing the Virasoro vacuum block.

\section{Applications and discussion}\label{v}

We have presented three representations of the four-point Virasoro conformal blocks on the sphere, for generic internal and external dimensions and central charge. Each of these is a partial resummation of a closed-form series expansion, obtained by solving recursion relations for the expansion coefficients. This formally eliminates the need for recursion. Optimistically, these representations may help us to eventually obtain a complete understanding of the functional properties of Virasoro blocks. 

We have not attempted to perform full resummations, although we hope that our expressions will lead to that. To this end, it would be quite desirable to massage our coefficients, which are somewhat difficult to work with, into simpler forms. It would be remarkable if, for completely generic parameters, the fully re-summed Virasoro blocks can be expressed solely in terms of known special functions. A related obvious goal for the future is to use our findings to derive new properties of the Virasoro blocks analytically. We will say more about this below. 

An orthogonal issue that deserves further attention is whether the representations herein are computationally more efficient to implement than the recursion procedure. This would enable faster numerical calculation of correlators and checks of crossing symmetry. Using a straightforward Mathematica code to automate \eqr{sol1} for example, preliminary investigations indicate that this generates the symbolic Virasoro blocks an order of magnitude faster than standard recursion algorithms; on the other hand, plugging in specific values for the parameters $(h_i,h_p,c)$ does not improve upon the standard algorithm (which is already quite efficient). 

It would be interesting to understand our results from the perspective of the AGT correspondence. For instance, the global representation of the Virasoro block seems not to have a clear interpretation from the 4d gauge theory point of view. 
 
In the remaining subsections, we discuss some applications of our results at large $c$ and directions for further work. 

\subsec{Semiclassical limits}

\subsubsection{Corrections to the heavy-light limit}

In \eqr{sci}--\eqr{sck}, we have presented the full set of $1/c$ corrections to the individual blocks, written in an expansion in $w$ that converges for $|w|<1$. These may be a good starting point for understanding thermal physics in AdS$_3$/CFT$_2$ beyond the leading heavy-light semiclassical limit \cite{fkw2}. An immediate question is how the subleading corrections to the heavy-light conformal blocks are able to resolve the microstate, and what these corrections can tell us about the departure from thermality, and/or late-time behavior of two-point correlators in black hole backgrounds. There may also be non-perturbative corrections to the blocks that are only visible at finite $w$.\footnote{Naively, it seems that individual blocks cannot receive non-perturbative corrections to the semiclassical limit because they are rational functions of $c$. This is too quick; but we note that if we write the corrections to the heavy-light semiclassical block at finite $w$ as
\e{}{\cF(c\rar\infty,h_i,h_p,z)\big|_{{h_H\over c}~\text{fixed}} \approx \cFsc(h_i,h_p,w) + {\cal G}(c,h_i,h_p,w)~,}
then ${\cal G}$, when expanded at small $w$ and large $c$, must only contain negative powers of $c$. This rules out an additive correction of the form ${\cal G} \sim e^{-c f(w)}$ where $f(w)\rar0$ for small $w$, for instance.} Once the corrections to the blocks are understood, one can try to understand to what extent the sum over blocks defining a given four-point function is approximately thermal in an expansion around large $c$. We leave this rich set of investigations for future work. 

\subsubsection{Exponentiation at large $c$}
Consider the following more traditional semiclassical limit, studied in the context of Liouville theory:
\e{llim}{c\rar\infty~, ~~h_i,h_p\rar\infty~, ~~ {h_i\over c}, {h_p\over c}~\text{fixed}~.}
It is believed that the Virasoro block exponentiates in this limit \cite{bpz, Zamolodchikov:426555, zamo2},
\e{sla}{\log \cF(c,h_i,h_p,z) \approx  -{c\over 6} f\left({h_i\over c}, {h_p\over c}, z\right) + O(c^0)~.}
There is significant evidence for this behavior, but it remains unproven. (See Appendix D of \cite{Harlow:2011ny} for a clear exposition of this idea.) Assuming exponentiation, $f$ is typically determined via a monodromy prescription \cite{zamo2}, but is not known in closed form for finite $h_i/c$ and $h_p/c$.

It would be very interesting to prove directly, using our closed-form expressions, that exponentiation occurs. One could also assume exponentiation, and try to derive $f$ by isolating the $O(c)$ part of $\cF$ in the limit \eqr{llim}. This would be equivalent to a solution for the accessory parameter in the $SL(2,\mathbb{C})$ monodromy problem, and to a non-series solution of the Painlev\'e VI equation subject to the boundary conditions put forth in  \cite{Litvinov:2013sxa}. 

Let us make one simple remark about the exponentiation of the vacuum block.\footnote{Deriving the exponentiated vacuum block, $f_{\vac}$, would have far-reaching consequences, including the determination of the two-interval R\'enyi entropy, $S_n$, for arbitrary interval spacings in the vacuum of sparse, large $c$ CFTs \cite{Hartman:2013mia}; the Einstein action evaluated on handlebody geometries asymptotic to replica manifolds \cite{Faulkner:2013yia}; and the extremal CFT partition functions on genus-$(n-1)$ replica manifolds at large $c$, and by extension, the Schottky problem for replica manifolds \cite{Yin:2007gv}.} From \eqr{cbj}--\eqr{cvac}, one deduces that in order for the exponentiation \eqr{sla} to occur, the coefficient of $z^K$, $\cF_{\vac,K}$, must not have terms of $O(c^{\lfloor K/2\rfloor+1})$ or greater in the limit. One can easily see\footnote{The terms in \eqr{44} are the contributions of operators made of $\lfloor K/2\rfloor$ stress tensors, which have norms scaling as $c^{\lfloor K/2\rfloor}$ and contribute $\lfloor K/2\rfloor$ powers of $h_i$ to each OPE coefficient.} that indeed, in any limit in which $h_i\rar\infty,c\rar\infty$,
\e{44}{\cF_{\vac,K}(h_i,z) \sim \left({h_1h_3\over c}\right)^{\lfloor K/2\rfloor} + \text{(subleading)}~~ \rar~~ c^{\lfloor K/2\rfloor} + O(c^{\lfloor K/2\rfloor-1})}
where the arrow implements the specific limit \eqr{llim}.  Furthermore, if one instead considers the ``Newtonian'' limit of large $c$ with ${h_1h_3\over c}$ fixed -- thereby neglecting the subleading terms in \eqr{44} -- the block has already been proven to exponentiate \cite{fkw1}. 
The key point here is that one should regard the Newtonian limit as capturing the leading terms in the limit \eqr{llim}. A re-summation of all subleading corrections to the Newtonian limit, or to the heavy-light limit, would give the full vacuum block in the limit \eqr{llim}; the heavy-light corrections can be systematically computed, at least in principle, using \eqr{scg}.

\subsec{Entanglement}

Let us derive some results that follow from existing literature on large $c$ Virasoro blocks.
\vs
We first consider two-interval R\'enyi entropy on the plane. This is defined by a four-point function of twist fields $\Phi_{\pm}$, 
\e{era}{S_n = {1\over 1-n} \log \langle \Phi_+(\infty) \Phi_-(1) \Phi_+(z) \Phi_-(0)\rangle_{\C^n/\Z_n}~.}
The correlator is evaluated in the cyclic orbifold CFT, denoted $\C^n/\Z_n$,  with central charge $c_n:= nc$. The cross ratio $z$ is real. The twist fields have holomorphic weight
\e{erb}{h_{\Phi} = {c_n\over 24}\left(1-{1\over n^2}\right)~.}
In a sparse large $c$ CFT, the vacuum block, which exponentiates as in \eqr{sla}, is believed to dominate the twist correlator at leading order \cite{Hartman:2013mia}. Including the anti-holomorphic vacuum block, the R\'enyi entropy is
\e{erc}{S_n \approx -{c_n\over 3(1-n)}f_{\vac}(z)}
where $f_{\vac}(z) \approx -{6\over c}\log \cF_{\vac}$. This is supposed to be valid for $0\leq z\leq \half$ where the $s$-channel is dominant, with a phase transition to the $t$-channel occurring at $z=\half$. 

Away from the entanglement entropy limit $n\rar 1$, and the case $n=2$ (for which $S_2$ is the free energy of the CFT on a torus), $S_n$ has only been computed in a short interval expansion, $z\ll 1$ \cite{Headrick:2010zt, Calabrese:2010he}. We can obtain $S_n$ for {\it finite} $z$ by instead expanding in $\d n := n-1$, as we now show. This expansion has been studied in higher dimensions to obtain R\'enyi entropies in interacting CFTs \cite{Perlmutter:2013gua, Lee:2014zaa, Hung:2014npa}.

To make contact with previous work on computing $f_{\vac}$, we use the light/heavy notation of \eqr{scb} with $\d_L=\d_H=0$, and define rescaled dimensions
\e{}{\eps_i := {6 \over c}h_i~.}
This notation is useful because $f_{\vac}$ is only known to linear order in small $\eps_L$ but for generic $\eps_H$ \cite{fkw1}. Expanded in small $\eps_H$, 
\es{erd}{f_{\vac}(z) &\approx \eps_L\Bigg[2\log z -{\eps_H\over 3}z^2\hyp(2,2,4;z) \\&+ \eps_H^2\left({4(z-1)\log^2(1-z)\over z^2}+\left({4\over z}-2\right)\log(1-z) + 8\right) + O(\eps_H^3)\Bigg]\\
&+ \eps_L^2\left[\eps_H\left({4(z-1)\log^2(1-z)\over z^2}+\left({4\over z}-2\right)\log(1-z) + 8\right) + O(\eps_H^2)\right]+O(\eps_L^3)}
We have inferred the $\eps_L^2\eps_H$ term from the invariance of $\cF_{\vac}$ under $h_L \leftrightarrow h_H$. The same logic leads to the full set of terms linear in $\eps_H$ for generic $\eps_L$. 

To compute $S_n$ for two intervals in the vacuum, we take all external operators to be twist operators, so $\eps_L = \eps_H = \eps_{\Phi}$. In the $\d n$ expansion,
\e{ere}{\eps_{\Phi} = {6\over c_n} h_{\Phi} = {1\over 4}\left(1-{1\over n^2}\right)\approx \half \d n - {3\over 4} \d n^2 + O(\d n^3)~.}
Thus, the $\eps$ expansion of $f_{\vac}$ is the $\d n$ expansion of the R\'enyi entropy at leading order in large $c$. Defining the mutual R\'enyi information, $I_n = -S_n +{c\over 6}\left(1+{1\over n}\right)\log z$, subtracts the leading log term. Putting everything together, 
\es{erf}{I_n(z)\big|_{O(c)} &= \d n\left({z^2\over 36}\, \hyp(2,2,4;z)\right) \\&+  \d n^2\left(\frac{z(2-z)\log(1-z)+2(1-z)\log^2(1-z)}{6z^2}\right)+O(\d n^3)~.}
Expanded at small $z$, this agrees with the short interval expansion, computed in \cite{Chen:2013dxa} through $O(z^9)$ for generic $n$. 

Note also that precisely the same analysis can be done for the ``spin-3 R\'enyi entropy'' of \cite{Hijano:2014sqa}, using the semiclassical $W_3$ vacuum block \cite{deBoer:2014sna}.  %
\vs
\vs
$f_{\vac}$ can also be co-opted to give the large $c$ R\'enyi entropy for a single interval in an excited state, non-perturbatively in the interval length but to linear order in $\d n$. In this case, $\eps_H$ and $\eps_L$ are distinct: $\Oc_L$ is a twist operator $\Phi_{\pm}$ in the $\eps_{\Phi}\rar 0$ limit, and $\Oc_H$ generates an excited state. The $\d n$ expansion corresponds to $\eps_{\Phi}\ll1$ for arbitrary $\eps_H$. From \eqr{erd}, we can only write down the explicit expression to $O(\d n)$ in an expansion in $\eps_H$:
\es{erg}{&S_n\big|_{O(c)} = S_{EE}\big|_{O(c)} \\
&+\d n \left[{\log z\over 6} -{\eps_H\over 12}\left({z^2\over 3}\, \hyp(2,2,4;z)+{4(z-1)\log^2(1-z)\over z^2}+\left({4\over z}-2\right)\log(1-z) + 8\right)\right]\\&+O(\d n^2,\eps_H^2)\nonumber}
where $z$ is a real interval length. This is valid in large $c$ CFTs in the range of $z$ for which the vacuum approximation is valid. $S_{EE}|_{O(c)}$ is computed\footnote{In our conventions, ${S_{EE}|_{O(c)} = \lim_{n\rar 1} {2\over 1-n}\log (z^{-2h_L}\cFsc_{\vac})}$, where $h_L = h_{\Phi}$.} from $\log \cFsc_{\vac}$ in \eqr{scc} or as a geodesic length in an asymptotically AdS$_3$ geometry \cite{Asplund:2014coa, Caputa:2014eta}. The remaining terms in the previous equation give a prediction for the on-shell Einstein action of a replica geometry with infinitesimal background charge $\eps_H$, expanded to $O(\d n, \eps_H)$. It would be interesting to verify this directly in the bulk. 

\vs \vs

The $1/c$ corrections to the heavy-light blocks give subleading corrections to the excited state entanglement entropy. Beyond leading order in large $c$, twist correlators are not dominated by the vacuum block for any finite range of $z$, even in sparse CFTs. This makes it difficult to make universal and concrete statements without further details about the CFT; however, the corrections \eqr{scj} apply to generic blocks, so our results may still be useful in this regard. This will be especially true if we can perform the sum over $w$ in \eqr{scj}.

\subsec{Scattering in AdS$_3$}
The $1/c$ expansion of Virasoro blocks contains information about the semiclassical expansion of scattering amplitudes in AdS$_3$ quantum gravity, with coupling constant $G_N = 3R_{\rm AdS}/2c$. Scattering processes are universally constrained by diffeomorphism invariance. This is the information contained in the Virasoro blocks. The details of the bulk spectrum and vertices map to the spectrum and OPE data of a dual CFT; these also obey constraints, such as the Cardy formula \cite{Cardy:1986ie}. 

In the large $c$ limit with all dimensions fixed, we wrote the explicit corrections to Virasoro blocks in \eqr{cm1}--\eqr{cm2}. These map to certain terms in the loop expansion of bulk four-point scattering of light particles. Unfortunately, it is hard to make a more precise statement: as is well-known from studies in higher dimensions, the map between bulk Witten diagrams and CFT global conformal blocks is not one-to-one (see e.g. \cite{Heemskerk:2009pn, ElShowk:2011ag, Fitzpatrick:2012cg}). One wonders whether the situation is simpler in AdS$_3$/CFT$_2$ thanks to Virasoro symmetry, but this is not obviously the case.\footnote{We thank Jared Kaplan for discussions on this subject.} As a concrete example, consider the $O(1/c)$ part of the vacuum Virasoro block, given in (2.34). The lone term is just the global block for stress tensor exchange. Perhaps contrary to expectations, this does not map directly to tree-level graviton exchange in the bulk. The Mellin amplitude for graviton exchange between massive scalar fields is, from Section 6.1.1 of \cite{Costa:2014kfa},
\e{}{M(s,t) = {2R_{AdS}\over c}\D_1\D_3\sum_{m=0}^{\infty}{Q_{2,m}(s)\over t-2m} + {3R_{AdS}\over c}{\Gamma(\D_1+\D_3-1)(s-\D_1\D_3-s(\D_1+\D_3))\over \Gamma^2(\D_1)\Gamma^2(\D_3)}}
with
\e{}{Q_{2,m}(s) =-\frac{3 s (2 m+s)}{2 m! (2)_m \Gamma (\D_1-m) \Gamma (\D_3-m)}~.}
$\D_i=2h_i$ are the scalar conformal dimensions given by $(mR_{AdS})^2 = \D(\D-2)$. Its transformation to position space does not simply yield a hypergeometric function.

One might wonder whether properties of graviton, rather than scalar, scattering may be more easily inferred from $1/c$ corrections to Virasoro blocks. The problem is that graviton bound states are composite operators dual to normal ordered products of the stress tensor, which is not a primary field. However, graviton scattering in AdS$_3$ is tree-level-exact when expressed in terms of the renormalized Newton constant, i.e. the central charge $c$, so the loop expansion is trivial anyway. In CFT terms, this is the statement that connected correlators of operators in the Virasoro vacuum module are proportional to $c$.\footnote{See \cite{Headrick:2015gba} for further justification of these and related statements about four-point scattering of spin-$s$ gauge fields in AdS$_3$.}

It would be very interesting to overcome these challenges and use \eqr{cm1}--\eqr{cm2} to make precise predictions for four-point scattering in AdS$_3$ quantum gravity.

\section*{Acknowledgments}

We thank Daniel Harlow, Jared Kaplan, Per Kraus, Sylvain Ribault, Grigory Tarnopolsky, Herman Verlinde and Ran Yacoby for helpful discussions. Support for this work comes from the Department of Energy under Grant No. DE-FG02-91ER40671.

\begin{appendix}

\sec{Properties of Virasoro blocks for pairwise identical operator dimensions}\label{AppB}

In this appendix we provide some analytic results on the behavior of the residues $R_{mn}$, and hence the series expansion coefficients $\gmn, \zeta_{mn}$ and $\gmn^{\sc}$. We will show that many of these vanish identically, especially in the case $h_p=0$, thereby making the evaluation of the full Virasoro block less computationally intensive than it first appears. 

For completely generic parameters $(h_i,h_p)$, none of the $\rmn$ vanishes. Henceforth we specialize to the pairwise limit, $h_1=h_2$, $h_3=h_4$, for generic $(h_1,h_3,c)$ unless stated otherwise. At times, we will suppress dependence on $h_i$ or $h_p$ to save space, but the pairwise property should be kept in mind. 

For pairwise blocks,
\e{apa}{P_{mn}(h_p) =  \prod_{j,k} \left(2\ell_1-{\ell_{jk}\over 2}\right)\left(2\ell_3-{\ell_{jk}\over 2}\right) \left({\ell_{jk}\over 2}\right)^2~.}
The ranges for $(j,k)$ were defined in \eqr{range}. We immediately note that $P_{mn}(h_1=h_2,h_3=h_4,h_p)=0$ for $mn$ odd, because $p=q=0$ is guaranteed to appear in the product, and $\ell_{00}=0$. When $h_p$ is generic, this cannot be cancelled by a corresponding pole in $A_{mn}$. Thus we conclude that
\e{apb}{R_{mn}(h_p)=0 ~~\text{for}~mn~\text{odd}~.}
This implies that the pairwise Virasoro block receives no quasi-primary contributions at all odd levels above the primary. In the global representation \eqr{sol1}, for instance, 
\e{apc}{\chi_q(h_p)=0~~\text{for}~q~\text{odd}~.}

\subsec{Vacuum block}
We now further specialize to $h_p=0$. This leads to extra simplifications that can be chalked up to the rationality of $t_{mn}(0)$:
\e{apd}{t_{mn}(0) ={m+\a\over n+\a}~,~~\text{where}~\a:= \sgn(n-m)~.}
The $\ell_{jk}$ now become
\e{ape}{\ell_{jk}(m,n,0) = {j(n+\a)-k(m+\a)\over \sqrt{(m+\a)(n+\a)}}~.}
Thus, $\ell_{jk}(m,n,0) = 0$ when $j(n+\a)=k(m+\a)$. 

We now show that for generic pairwise identical operator dimensions, $R_{mn}$ obeys the properties \eqr{three}--\eqr{third} in the main text:
\vs
{\bf i)} The following argument proves the validity of {\bf i)}, which is the vacuum limit of \eqr{apb}.\footnote{As a direct calculation, this is more subtle than the $h_p\neq 0$ case because of the rationality of $t_{mn}(0)$, and hence the presence of extra poles in $\amn$. Indeed, there are cases such as $(m,n)=(3,5)$ where $\amn$ has the same number of poles as $\pmn$ has zeroes. The argument below guarantees that the correct way to take the limit of their product gives zero.} As discussed around \eqr{chiq}, the coefficient $\chi_q$ computes a sum over squared OPE coefficients, divided by the norms of the internal quasi-primary. In taking the vacuum limit $h_p\rar 0$ of the general result, one cannot encounter divergences. So for pairwise blocks, $\chi_q=0$ for odd $q$, for all $h_p$. This implies $\rmn=0$ for all $mn$ odd and all $h_p$, because $\chi_q$ is a linear combination of products of $\rmn$, and there can be no cancellations thanks to the $h_i$ dependence. Therefore, 
\e{apj}{R_{mn}(0)=0 ~~\text{for}~mn~\text{odd}~,}
and the vacuum Virasoro block receives no quasi-primary contributions at all odd levels above the primary. This can be checked experimentally.

\vs
{\bf ii)} Recalling from \eqr{range} that $j_{max} = m-1$ and $k_{max} = n-1$, a double zero in \eqr{apa} is guaranteed to occur when $\a=-1$, that is, when $m>n$. Therefore, $P_{mn}(h_1=h_2, h_3=h_4,0)=0$ when $m>n$. On the other hand, $A_{mn}$ also develops a simple pole at this same point. If there were no other zeroes allowed in the products that define $\pmn$ and $\amn$, then $R_{mn}$ would vanish for $m>n$. But this is not the case in general, because $A_{mn}$ actually can have more poles. We content ourselves with experimental data on their multiplicities. For all pairs $(m,n)$ where $m\geq 1, n\geq 2$ up to level $mn=30$, we find that indeed,
\e{apfa}{R_{mn}(0) = 0 ~~\text{for}~~m>n~.}
We believe this to hold in general. 

In the special case $m=n$, we have $t_{mm}(0) = 1$, which implies
\e{aph}{\ell_{jk}(m,m,0) = {j-k}~.}
By inspecting the number of times that $j=k$ and using the definition of $R_{mm}$, one can quickly prove that $R_{mm}$ vanishes quadratically in $\ell_{jj}$. We have also checked this explicitly through $m=9$. Therefore,
\e{api}{R_{mm}(0) = 0~.}
\vs
{\bf iii)} Finally, let us consider the case when one pair of external operators is the identity, $h_1=h_2=0$.  The four-point function reduces to a two-point function, and in particular,
\es{apl}{\lim_{h\rar 0}C_{hhh_p} = 
\left\{ 
\begin{array}{cc}
1 \, , & h_p=0 \\ 
0 \,, & ~h_p\neq 0 \,,
\end{array}
\right.
}
The top line implies that the conformal block must trivialize for $h_p=0$, which in turn implies
\e{apm}{\rmn(h_1=0,h_3,h_p=0) = 0~~\text{for all}~(m,n)~.}

This is straightforward to confirm by direct calculation. Because the $h_1$ dependence of $\rmn$ is in $P_{mn}$ alone, specifically in the terms $2\ell_1-\ell_{jk}/2$ in \eqr{apa}, we must show that this vanishes for some allowed $(j,k)$. In the limit $h_1=0$, 
\e{apn}{2\ell_1(m,n,h_1=0,0) = \ell_{11}(m,n,0) = {1-t_{mn}(0)\over \sqrt{t_{mn}(0)}}~.}
Using \eqr{apd}, one then has, for $m<n$, 
\e{}{ \ell_{11}(m,n,0) - \half\ell_{jk}(m,n,0) = {-j(n+1)+k(m+1)+2(n-m)\over \sqrt{(m+1)(n+1)}}~.}
This vanishes for $(j,k) = (j_{min},k_{min}) =(-m+1,-n+1)$. 

Note that the property \eqr{apm} does not hold for $h_p\neq0$, because the OPE coefficient multiplying the block vanishes, $C_{hhh_p}=0$; this guarantees the triviality of the four-point function, but the calculation of the conformal block itself need not, and in fact does not, trivialize.\footnote{This can be checked using a brute force calculation of the conformal block up to level two: for $h_1=h_2=0$ and $h_3=h_4$ arbitrary, one finds
\e{}{\cF_2 = \frac{h_p \left(c (h_p+1)^2+8 h_p \left(h_p^2+h_p-1\right)+4 (h_p-1) h_3\right)}{c (8 h_p+4)+8 h_p (8 h_p-5)}}}

\sec{Vacuum module data}\label{AppA}
We list the quasi-primaries $O$ in the vacuum Verma module of the Virasoro algebra through level 6, as well as their norms and OPE coefficients. For $O$ of weight $H$ admitting a mode expansion 
\e{}{O(z) = \sum_{n\in\mathbb{Z}}{O_n\over z^{n+H}}~,}
its norm, $\N_O$, and normalized OPE coefficient with a primary $\Oc$ of weight $h$, $C'_{hhO}$, are defined as
\es{}{\N_{O} = \langle O_H O_{-H}\rangle~, \quad C'_{hhO} = {\langle \Oc_h \Oc_{-h+H} O_{-H}\rangle\over \N_{h}}~.}
The state-operator correspondence is $O(0) = O_{-H}|0\rangle$; we will use a shorthand $O=O_{-H}$.

At level 2, we have the stress tensor, $T = L_{-2}$, which has 
\e{}{\N_T = {c\over 2}~, \quad C'_{hhT} = h~.}

At level 4, we have the operator $\L = L_{-2}^2-\frac35L_{-4}$, the normal ordered product of two $T$'s with a derivative term removed, which has 
\e{}{\N_{\L} = {c(5c+22)\over 10}~, \quad C'_{hh\L}=   h^2+{h\over 5}~.}

At level 6, we have two quasi-primaries
\es{}{\cO^{(1)}&=-\frac{20}{63}L_{-6}-\frac{8}{9}L_{-4}L_{-2}+\frac{5}{9}L_{-3}L_{-3},\\
\cO^{(2)}&=-\frac{60c+78}{70c+29}L_{-6}-\frac{3(42c+67)}{70c+29}L_{-4}L_{-2}+\frac{93}{70c+29}L_{-3}L_{-3}+L_{-2}L_{-2}L_{-2}} 
which have
\e{}{\N_{\Oc^{(1)}} =  \frac 4{63}c\left(70c+29\right)~, \quad C'_{hh\Oc^{(1)}}  = -{2\over 63} h\,(14h+1)}
and
\e{}{\N_{\Oc^{(2)}} = \frac34\,\frac{c\,(2 c-1)\,(5c+22)\,(7c+68)}{70c+29} ~, ~ C'_{hh\Oc^{(2)}}  = \frac{h \left(c \left(70 h^2+42 h+8\right)+29 h^2-57 h-2\right)}{70 c+29}}

\section{Explicit series expansions}\label{appc}
For reference, we expand the global and semiclassical representations of the Virasoro block through level six. 

\subsection*{Global representation}

Suppressing the $h_i$ dependence (which does not participate in the algorithm) to save space, \eqr{sol1} yields 
\es{ex}{&\chi_2(c,h_p)=\g_{12}(c,h_p)\\
&\chi_3(c,h_p) = \g_{13}(c,h_p)\\
&\chi_4(c,h_p) = \g_{14}(c,h_p) + \g_{22}(c,h_p) + \g_{12}(c,h_p)\cdot \g_{12}(c_{12}(h_p), h_p+2)\\
&\chi_5(c,h_p) = \g_{15}(c,h_p) + \g_{12}(c,h_p)\cdot\g_{13}(c_{12}(h_p), h_p+2) + \g_{13}(c,h_p)\cdot\g_{12}(c_{13}(h_p), h_p+3)  \\
&\chi_6(c,h_p)=\g_{16}(c,h_p) + \g_{23}(c,h_p) + \g_{32}(c,h_p) \\
&\quad\quad\quad~~ + \g_{12}(c,h_p)\cdot\g_{14}(c_{12}(h_p), h_p+2) + \g_{12}(c,h_p)\cdot\g_{22}(c_{12}(h_p),h_p+2)\\
&\quad\quad\quad~~ + \g_{14}(c,h_p) \cdot \g_{12}(c_{14}(h_p), h_p+4) + \g_{22}(c,h_p)\cdot \g_{12}(c_{22}(h_p), h_p+4)\\
&\quad\quad\quad~~ +\g_{13}(c,h_p)\cdot \g_{13}(c_{13}(h_p),h_p+3)\\
&\quad\quad\quad~~ +\g_{12}(c,h_p)\cdot\g_{12}(c_{12}(h_p),h_p+2)\cdot\g_{12}(c_{12}(h_p+2),h_p+4)}
It is straightforward to reproduce this by iterating \eqr{rec2}. If two external operators have identical dimension, \eqr{resv} tells us that $\g_{mn}=0$ for $mn$ odd. If we further specify $h_p=0$, \eqr{three} tells us that all $\g_{m_1n_1}(c,0)$ with $m_1\geq n_1$ also vanish. Using a computer, one can easily check that \eqr{ex} reproduces the OPE results \eqr{cvac}.

\subsection*{Vacuum block in the semiclassical representation}\label{appd}

We focus on the vacuum block coefficients, $\csc_{\vac,q}$. We trade $\gsc$ for $\gmn$ using their definition \eqr{scf},
\e{}{\gsc(c,h_i,h_p) = \gmn(c,\lbrace h_L,\eta c_{mn}(h_p)\rbrace,h_p)~.}
Then \eqr{scg} yields
\es{}{\chi^{sc}_{\vac, 2}(c,h_i) &=0\\
\chi^{sc}_{\vac,4} (c,h_i)&= \g_{14}(c, \lbrace h_L, \eta c_{14}(0)\rbrace,0)\\
\chi^{sc}_{\vac, 6}(c,h_i) &= \g_{16}(c, \lbrace h_L, \eta c_{16}(0)\rbrace,0) + \g_{23}(c, \lbrace h_L, \eta c_{23}(0)\rbrace,0) \\&+ \g_{14}(c, \lbrace h_L, \eta c_{14}(0)\rbrace,0)\cdot \g_{12}(c_{14}(0), \lbrace h_L, \eta c_{12}(4)\rbrace,4)}
where
\es{}{  \g_{14}(c, \lbrace h_L, \eta c_{14}(0)\rbrace,0)&= \frac{2 \eta (1-22 \eta) h_L (5 h_L+1) }{5 (5 c+22)}\\
\g_{16}(c, \lbrace h_L, \eta c_{16}(0)\rbrace,0) &= \frac{4 \eta (1-34 \eta) (68 \eta-3) h_L (7 h_L+2) (7 h_L+3) }{3003 (7 c+68)}\\
\g_{23}(c, \lbrace h_L, \eta c_{23}(0)\rbrace,0) &= \frac{(\eta-1) \eta (8 \eta-1) h_L (2 h_L (16 h_L-9)+1) }{21021 (2 c-1)}\\
\g_{12}(c_{14}(0), \lbrace h_L, \eta c_{12}(4)\rbrace,4)&=\frac{10}{441} (1-36 \eta) (3 h_L+2)}
Plugging into \eqr{qp2sc} and writing $(\eta,h_L) =(h_1/ c, h_3)$, this matches $\cF_{\vac}$ in \eqr{cbj} upon expanding $w(z)$ around $z=0$. As discussed below, $\csc_{\vac, q}$ requires computation of fewer terms then $\chi_{\vac, q}$ does, which is clear upon comparing to \eqr{ex}. 

\vs

We now want to answer the question of how many of the terms in the sum \eqr{scg} vanish identically due to the fact that $c_{12}(0)=0$, as discussed at the end of Section \ref{iii}. A lower bound can easily be obtained by mapping the problem to one of counting compositions (i.e. ordered partitions) of $2j$ into positive even integers, which we call $p_{\ell}$.

Let's first establish how to exactly count the number of terms appearing in the naive sum before accounting for $c_{12}(0)=0$. As explained in the text and in Appendix \ref{AppB}, when $h_p=0$ the ranges for the $(m_{\ell}, n_{\ell})$ are $m_{\ell}\geq 1$ and $n_{\ell}\geq 2$, with the additional conditions that $m_1<n_1$ and that $m_{\ell}n_{\ell}$ be even. This gives us the algorithm for counting terms in the sum: first, we count compositions of $2j$ into positive even integers $p_{\ell}:= m_{\ell}n_{\ell}$; then, we count multiplicative partitions of $p_{\ell}$ into $(m_{\ell},n_{\ell})$, subject to the constraints on the ranges of $m_{\ell}$ and $n_{\ell}$ just quoted. 

The first step is elementary. The number of compositions of $2j$ into positive even integers is the same as the number of compositions of $j$ into positive integers. This number is $2^{j-1}$. We now want to ask how many of these terms vanish identically due to the exclusion $p_1\neq2$, which is the statement that $c_{12}(0)=0$. This halves the number of compositions: of the $2^{j-1}$ total compositions, $2^{j-2}$ terms have $p_1=2$, which are the terms that vanish identically. Therefore, the restriction $m_1n_1\neq 2$ associates half of the compositions to terms that vanish.

Considering the second step of the algorithm, one easily sees that slightly more than half of the terms in the sum \eqr{scg} vanish, due to the extra restriction that $m_1<n_1$. This was the claim in the text. 

\end{appendix}

\bibliographystyle{ssg}
\bibliography{biblio}

\end{document}